\documentclass[aps,reprint,superscriptaddress,nofootinbib]{revtex4-1}

\usepackage{graphicx}
\usepackage{amssymb}
\usepackage{mathrsfs}
\usepackage{color}
\usepackage{natbib}
\usepackage{footnote}

\begin{document}

\title{Discovery of a new extragalactic population of energetic particles.}

\author{Anthony M. Brown}
\affiliation{Department of Physics and Centre for Advanced Instrumentation, University of Durham, South Road, Durham, DH1 3LE, UK}
\author{C\'eline B\oe hm}
\affiliation{Institute for Particle Physics Phenomenology (IPPP), University of Durham, Durham DH1 3LE, UK}
\affiliation{LAPTH, U.~de Savoie, CNRS,  BP 110, 74941 Annecy-Le-Vieux, France}
\author{Jamie Graham}
\affiliation{Department of Physics and Centre for Advanced Instrumentation, University of Durham, South Road, Durham, DH1 3LE, UK}
\author{Thomas Lacroix}
\affiliation{Institut d'Astrophysique de Paris, UMR 7095, CNRS, UPMC Universit\'e Paris 6, Sorbonne Universit\'e, 98 bis boulevard Arago, 75014 Paris, France}
\author{Paula Chadwick}
\affiliation{Department of Physics and Centre for Advanced Instrumentation, University of Durham, South Road, Durham, DH1 3LE, UK}
\author{Joseph Silk}
\affiliation{Institut d'Astrophysique de Paris, UMR 7095, CNRS, UPMC Universit\'e Paris 6, Sorbonne Universit\'e, 98 bis boulevard Arago, 75014 Paris, France}
\affiliation{Department of Physics \& Astronomy, Johns Hopkins University, Baltimore, MD 21218, USA}

\date{\today}

\begin{abstract}
We report the discovery of a statistically significant hardening in the \textit{Fermi}-LAT $\gamma$-ray spectrum of Centaurus A's core, with the spectral index hardening from $\Gamma_{1}=2.73 \pm 0.02$  to $\Gamma_{1}=2.29 \pm 0.07$ at a break energy of ($2.6 \pm 0.3$) GeV. Using a likelihood analysis, we find no evidence for flux variability in Cen A's core lightcurve above or below the spectral break when considering the entire 8 year period. Interestingly, however, the first $\sim3.5$ years of the low energy lightcurve shows evidence of flux variability at the $\sim3.5 \sigma$ confidence level. To understand the origin of this spectral break, we assume that the low energy component below the break feature originates from leptons in Centaurus A's radio jet and we investigate the possibility that the high energy component above the spectral break is due to an additional source of very high energy particles near the core of Cen A. We show for the first time that the observed $\gamma$-ray spectrum of an Active Galactic Nucleus is compatible with either a very large localized enhancement (referred to as a spike) in the dark matter halo profile or a population of millisecond pulsars. Our work constitutes the first robust indication that new $\gamma$-ray production mechanisms can explain the emission from active galaxies and could provide tantalizing first evidence for the clustering of heavy dark matter particles around black holes.
\end{abstract}

\pacs{}

\maketitle


\section{Introduction}
At a distance of 3.7 Mpc, Centaurus A (Cen A) is the closest known $\gamma$-ray emitting Active Galaxy (e.g. \cite{ferr}). With a Fanaroff-Riley type I (FR I) radio morphology, Cen A possesses a relativistic jet that is orientated at a large angle relative to Earth's line of sight. As such, Cen A is believed to belong to the parent population of BL Lac objects within the unified model of active galactic nuclei (AGN; \cite{fr,urry}). The off-axis nature of Cen A's jet allows for an easy decoupling of the observed relativistic effects from the intrinsic properties of the jet. Furthermore, since the jet emission is weakly beamed, different emission components which typically are not observed in BL Lacs due to the dominant beamed jet component, may become visible. 

Due to its proximity, Cen A affords us an excellent opportunity to study the physics of relativistic outflows. As such, Cen A has been extensively studied at many wavelengths. Radio observations have found Cen A to possess a variety of radio structures, with size scales ranging from arcseconds in the inner jet to giant radio lobes which extend $10^{\circ}$ across the sky. At X-ray energies, emission has been resolved from the inner jet which appears to be well collimated \cite{hardcastle,evans}. At $\gamma$-ray wavelengths, the first evidence for emission above 1 MeV was provided by balloon-borne experiments \cite{balloons}, with further evidence from the \textit{GRANAT} satellite \cite{granat}. Cen A was confirmed as a $\gamma$-ray point source by the OSSE and COMPTEL detectors onboard the \textit{Compton Gamma-Ray Observatory} in the $0.1-30$ MeV energy range, with the EGRET detector also discovering Cen A to be a source of $\gamma$-rays up to GeV energies \cite{steinle,Sreekumar1999}. At very high energies (VHE; $>250$ GeV), Cen A was discovered by the H.E.S.S. telescope array with a soft spectral index ($\Gamma$) of $-2.73 \pm 0.45_{\rm{stat}} \pm 0.2_{\rm{sys}}$ and a flux at 0.8\% of the Crab nebula \cite{HESSCenA2009}. 

The successful launch of the \textit{Fermi $\gamma$-ray Space Telescope}, and the unprecedented sensitivity and resolution of the Large Area Telescope (LAT) onboard \textit{Fermi}, affords us an ideal opportunity to understand the inner working of Cen A. The first ten months of LAT observations confirmed the core of Cen A to be a source of MeV and GeV $\gamma$-rays \cite{latcore}, with  $\Gamma = -2.69 \pm 0.10_{\rm{stat}} \pm0.08_{\rm{sys}}$ and a flux of ($1.50 \pm 0.25_{\rm{stat}} \pm 0.37_{\rm{sys}}$)$\times10^{-7}$ ph cm$^{-2}$~s$^{-1}$. A deeper 4-year analysis of Cen A's core with \textsc{pass7} data revealed a similar spectral index and flux, with the authors noting that the spectrum above $\sim4$ GeV appeared to depart from a power-law \cite{saha2013}. This departure was statistically insignificant, with a broken power-law preferred over a power-law at the $<3\sigma$ level. Interestingly, early \textit{Fermi}-LAT observations revealed extended $\gamma$-ray emission spatially coincident with Cen A's giant radio lobes \cite{latlobe}. Both core and lobe emissions appeared to have similar luminosities, and neither showed variability in their flux. The lack of variability in Cen A's flux is at odds with properties of other prominent $\gamma$-ray bright radio galaxies, such as M87 and NGC 1275, whose flux varies on very short timescales (as little as days) \cite{Aharonian2006b,Brown2011}. 

Building on earlier work that investigated the contribution to M87's $\gamma$-ray emission from dark matter annihilation, \cite{M87_limits_my_paper}, in this paper we investigate the spectral and temporal properties of the $\gamma$-ray emission from Cen A's core. Using 8 years of \textit{Fermi}-LAT data, and taking advantage of the improvements in effective area, energy and angular resolution afforded by the \textsc{pass8} data analysis, we discover a statistically significant hardening feature in Cen A's MeV-GeV spectrum. 

The structure of our paper is as follows. After establishing in Section II the statistical significance of the spectral hardening feature, we investigate the temporal characteristics of the flux below and above the onset of the spectral feature. In Section III we investigate two possible interpretations; one involves the annihilation products from dark matter with a `spike' density profile and the other one millisecond pulsars. 

\section{\textit{Fermi}-LAT observations and data analysis \label{sec:analysis}}

The LAT detector onboard \textit{Fermi}, described in detail in \cite{lat,lat2}, is a pair-conversion telescope observing a photon energy range from below 20 MeV to above 500 GeV, with unprecedented sensitivity and resolution compared to previous $\gamma$-ray space missions. The \textsc{pass8} data analysis and instrument response functions (IRFs) of the LAT detector has brought further improvement in detector performance. Since 2008 August 4, the vast majority of data taken by \textit{Fermi} has been in all-sky-survey mode. This observing mode, coupled with the large effective area of the LAT detector and the long mission lifetime of the \textit{Fermi} mission, has produced the deepest extragalactic scan ever at MeV$-$GeV energies. This enables us to study the $\gamma$-ray emission from Cen A's core with unprecedented sensitivity and accuracy. 

The data used in this study comprise of all \textit{Fermi}-LAT event and spacecraft data taken during the first 8 years of the \textit{Fermi}-LAT science mission, from 2008 August 4 to 2016 August 8, equating to a Mission Elapsed Time (MET) period of 239557417 [s] to 492328870 [s]. All $0.1<E_{\gamma}<300$ GeV \textsc{source} events, across all point spread function (\textsc{psf}) classes, within a $15^{\circ}$ radius of interest (RoI) centred on the Cen A core position were considered. In accordance with criteria for \textsc{pass8} data analysis, a zenith cut of $90^{\circ}$ was applied, and good time intervals selected by removing data that did not satisfy the criteria `\textsc{DATA\_QUAL}$>0$ \&\& \textsc{LAT\_CONFIG}$==1$'. A summary of the criteria used in the analysis is given in Table 1. 

\begin{table}[h!]
\centering
   \caption{Summary of the criteria utilised in this analysis. Note that since a summed likelihood analysis was used, with \textsc{psf} classes 0 through 3 considered, four isotropic diffuse models were used.}
     \begin{tabular}{ll} \hline \hline 
      Science Tools version     & \textsc{v10r0p5}  \\ 
      IRF                       & \textsc{p8\_source\_v6}    \\ 
      Event class               & \textsc{source}, Pass 8     \\
      Photon Energies           & $0.1-300$ GeV    \\ 
      RoI                       & $15^{\circ}$      \\
      Zenith angle cut          & $<90^{\circ}$    \\  
      Data quality              & $>0$ \\
      LAT config                & $1$   \\
      Galactic diffuse model    & gll\_iem\_v06.fit \\
      Isotropic diffuse model   & iso\_P8R2\_SOURCE\_V6\_PSF0$-$3.txt \\ \hline \hline
    \end{tabular}
  \label{analysis}
\end{table}

The model employed during our likelihood analyses consisted of point-like $\gamma$-ray sources, spatially extended $\gamma$-ray sources and diffuse $\gamma$-ray emission. In particular, the position and spectral shape of all $\gamma$-ray point sources within $25^{\circ}$ of Cen A's core were taken from the Third \textit{Fermi} Source Catalog (3FGL; \cite{3fgl}). The extended sources considered were the two pulsar wind nebulae located within the $15-25^{\circ}$ annulus from Cen A's core, namely HESS J1303-631 and MSH 15-52, and Cen A's radio lobes \cite{latlobe}. The diffuse $\gamma$-ray emission detected by the LAT compresses of two components: the Galactic diffuse emission and the isotropic diffuse emission. The Galactic component of the diffuse emission was modelled with \textit{Fermi}'s gll\_iem\_v06.fit model. Since a summed likelihood analysis was used to take advantage of the performance of \textsc{pass8}'s four \textsc{psf}, the isotropic diffuse emission was defined by the iso\_P8R2\_SOURCE\_V6\_PSF0$-$3.txt files, where $0-3$ refers to \textsc{psf} class 0 through to 3. 

\begin{figure}[h!]
\includegraphics[width=1.0\columnwidth]{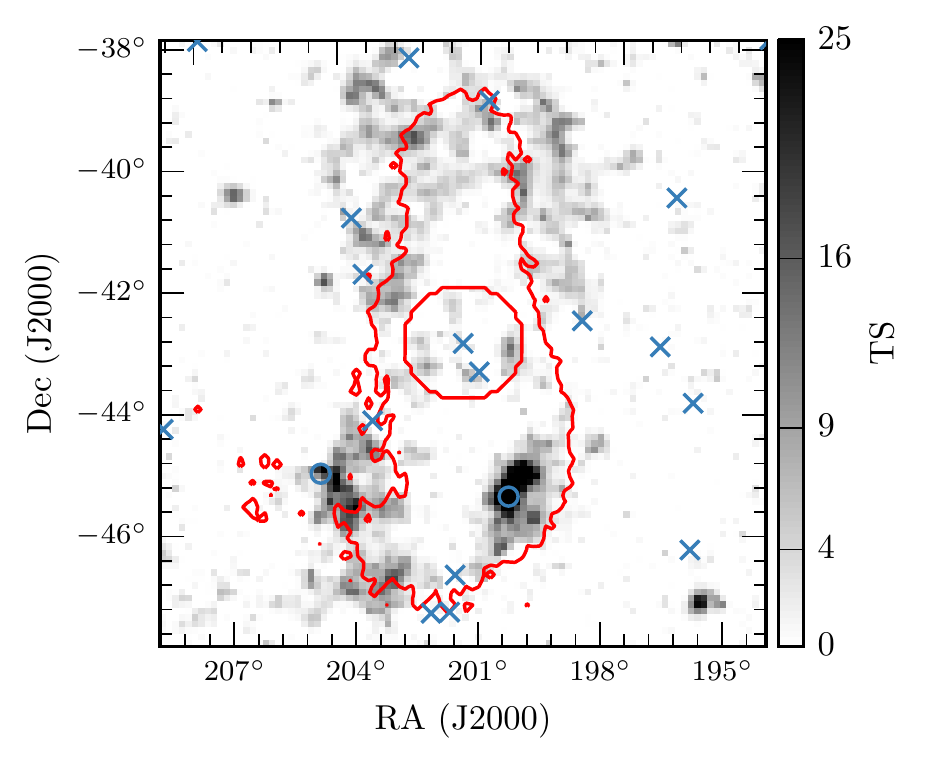}
\caption{A $17^{\circ} \times 17^{\circ}$ TS map of all $0.1-300$~GeV photons that passed the selection criteria, for the entire 8 year data set. The colour scale is in units of TS. Three new point sources were found, we note however that two of these apparent point sources are located on the boundary of Cen A's lobes, shown by the red contour lines.}
\label{allresids}
\end{figure}

To confirm the accuracy of our `diffuse $+$ point $+$ extended' model description, an initial \textsc{binned} likelihood analysis was performed over the entire dataset. Firstly, a likelihood analysis was performed with the normalisation of all sources within $15^{\circ}$ being left free to vary. Thereafter, a second likelihood fit was performed with all point sources with a test statistic\footnote{The test statistic, TS, is defined as twice the difference between the log-likelihood of two different models, $2[log L - log L_{0}]$, where $L$ and $L_0$ are defined as the likelihoods of individual model fits \cite{like}.}, TS, greater than 25 able to vary spectrally. Finally, all sources with $TS<1$ were removed from our model, and all sources with a 3FGL $TS_{var} > 72.44$ were fitted again with the normalisation and spectral shape left free to vary. The \textit{Fermi} science tool \textsc{gttsmap} was then used in conjunction with the final best-fit model from the three-step initial analysis, to construct a $17^{\circ} \times 17^{\circ}$ TS map centred on Cen A. This TS map was used to identify additional sources of $\gamma$-rays that were not accounted for in our best-fit model. As can be seen in Figure \ref{allresids}, there are three $>5\sigma$ excesses in the TS map indicating three new point sources of $\gamma$-rays have been found in our dataset. These new point sources were accounted for by a power-law fixed at the ($\alpha_{J2000}$,$\beta_{J2000}$) of the peak excess, and a final likelihood fit was performed with the normalisation and spectral index of the new point sources free to vary. Interestingly, we note that two of these point sources are located on the boundary of Cen A's lobes.

\begin{figure}[h!]
\includegraphics[width=1.0\columnwidth]{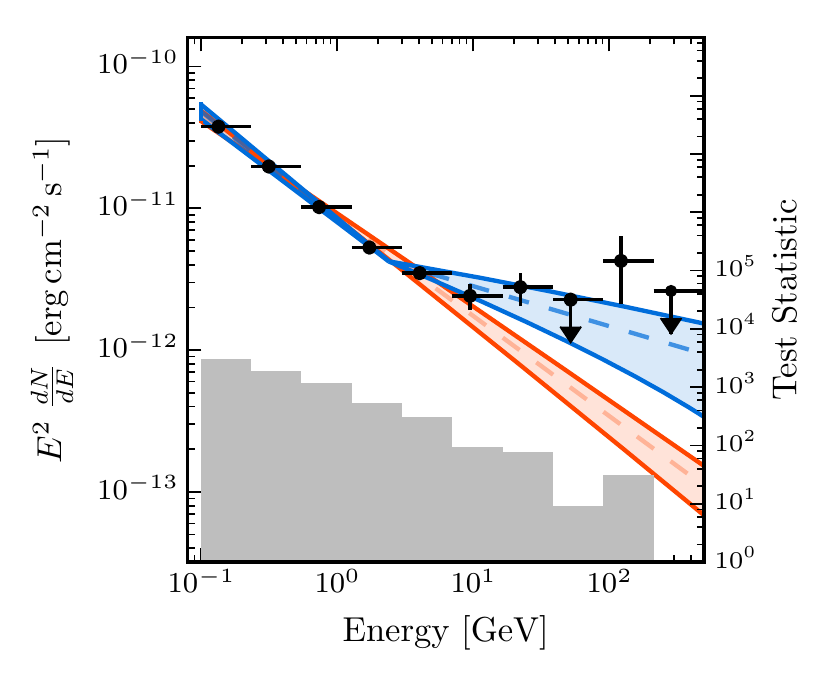}
\caption{The 0.1 to 300 GeV spectrum of Cen A's core, as seen by the \textit{Fermi}-LAT detector. The broken power-law model, shown in blue, is preferred over the power law model, shown in red, with a significance $>5\sigma$. The grey histogram shows the TS value for each spectral bin. The bins with a TS$<25$, are replaced with an upper limit at 95$\%$ confidence level.}
\label{spectrum}
\end{figure}

Once the dataset was modelled correctly, a summed-likelihood analysis was used to study the spectrum of Cen A's core. The data were binned into ten logarithmically spaced energy bins, with a likelihood fit being applied to each bin separately. For each separate likelihood fit, all spectral parameters were frozen except for the normalisation of Cen A's core. For an individual spectral bin, if the calculated flux had a $TS \ll 25$, a $2\sigma$ upper limit was calculated. The resulting spectrum can be seen in Figure \ref{spectrum}, with all error bars on the LAT data points representing a $1\sigma$ level of statistical uncertainty. The derived Cen A core spectrum seen in Figure \ref{spectrum} shows a clear departure from a simple power-law description above photon energies of $\sim1$ GeV. To quantify the significance of this discrepancy at high energies, a summed-likelihood analysis of the ($0.1-300$) GeV spectrum was undertaken with the Cen A's core being described by a broken power-law. A likelihood-ratio test of the broken power-law and power-law fit to Cen A core's ($0.1-300$) GeV spectrum found that the broken power-law spectral model is preferred over the power-law spectral model with a test statistic of $TS=28.6$, which equates to a significance slightly greater than $5\sigma$ (even after the increased number of free parameters of the broken power law). As such, this analysis represents the first observation of a statistically significant hardening of an AGN's $\gamma$-ray spectrum above a spectral feature.

From the best-fit broken power-law, the spectral break occurs at a photon energy of $E_{\rm{break}}= 2.6 \pm 0.3$ GeV, with a spectral index of $\Gamma_1 = -2.73 \pm 0.02$ below the spectral break, hardening to a spectral index of $\Gamma_2 = -2.29 \pm 0.07$ above the break energy. The total energy flux from Cen A's core in the $0.1-300$ GeV energy range is $\sim1.1\times10^{-12}$ ergs~cm$^{-2}$~s$^{-1}$. Assuming isotropic emission for the $\gamma$-ray flux and a luminosity distance of 3.7 million parsecs, the total luminosity of Cen A's core in the $0.1-300$ GeV is $1.8\times10^{39}$ ergs~s$^{-1}$. Interestingly, if we deconvolve the two spectral components, we find that the luminosity of the low and high energy components are comparable, with $9.9\times10^{38}$ ergs~s$^{-1}$ and $8.4\times10^{38}$ ergs~s$^{-1}$ respectively. 

\begin{figure}[h!]
\includegraphics[width=1.1\columnwidth]{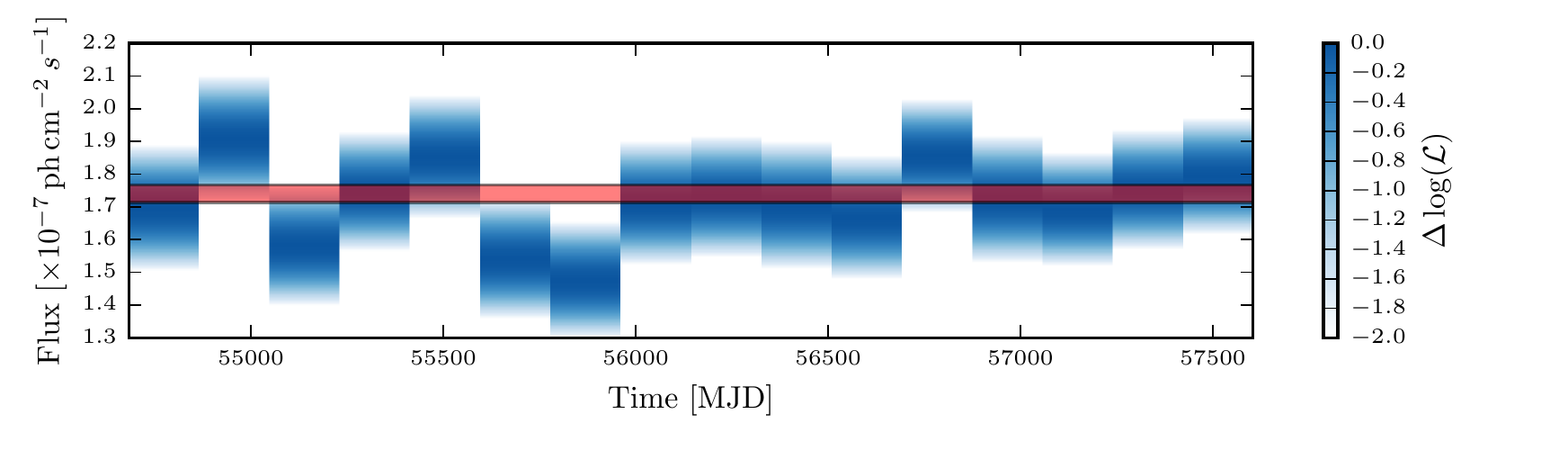}
\includegraphics[width=1.1\columnwidth]{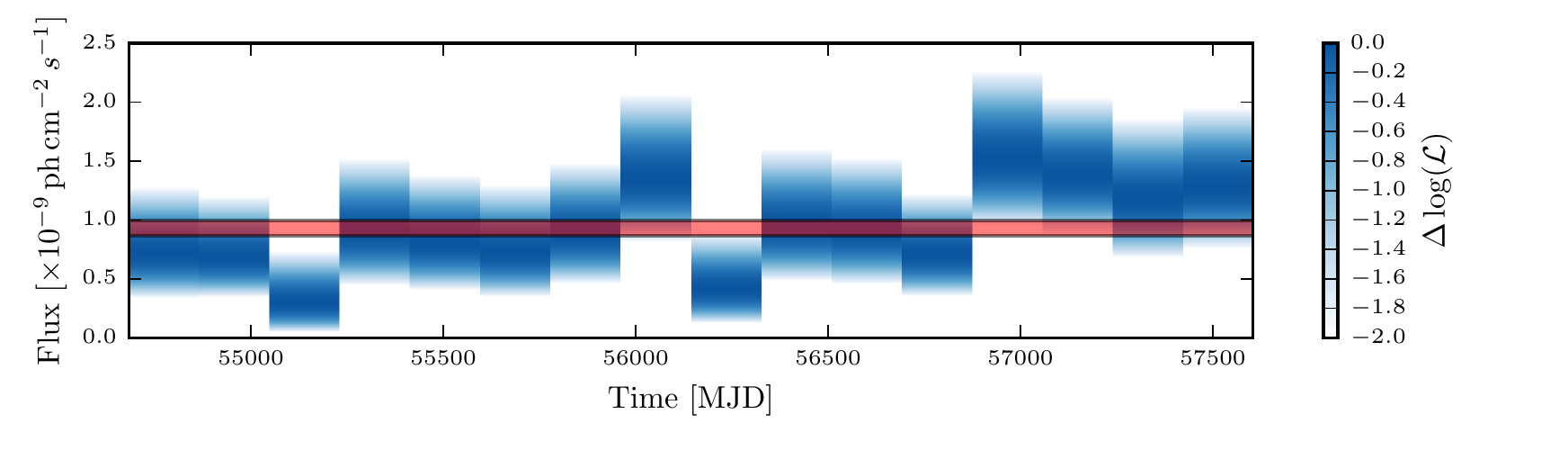}
\caption{\textit{Above}: light curve of $0.1-2.6$ GeV flux \textit{Below}: light curve of $2.6-300$ GeV flux, binned in six month temporal bins. The red horizontal band shows the 8-year averaged flux for each energy range.}
\label{lightcurves}
\end{figure}

To investigate the temporal characteristics of the two spectral components, we constructed individual lightcurves for the $\gamma$-ray flux above and below the break energy. To maintain significant statistics for each temporal bin in the high-energy lightcurve, the 8-year data set was binned into 6-month temporal bins. Using the global best-fit model with the normalisation of Cen A's core left free, a likelihood analysis was applied to each bin separately. The resulting lightcurves can be seen in Figure \ref{lightcurves}. 

To determine if there was evidence for flux variability in the resultant lightcurves, both a $\chi^2$ fit of the lightcurves to a constant flux value and a full likelihood evaluation, TS$_{\rm{var}}$ \cite{2fgl}, were used. When considering the 8-year data set, neither test found significant evidence of variability in either the low or high energy lightcurve. It is interesting to note that when considering the first 3.5 years of the low-energy lightcurve, there is evidence of variability at the $3.3\sigma$ level, with the remaining 4.5 years of the lightcurve being consistent with a constant flux. As such, while the flux above the spectral break is statistically consistent with being constant, there is evidence of variability in the low-energy spectral component during some of the time period we studied Cen A. 

\section{Interpretation  \label{sec:interpretation}}

\begin{figure*}
\includegraphics[width=0.8\linewidth]{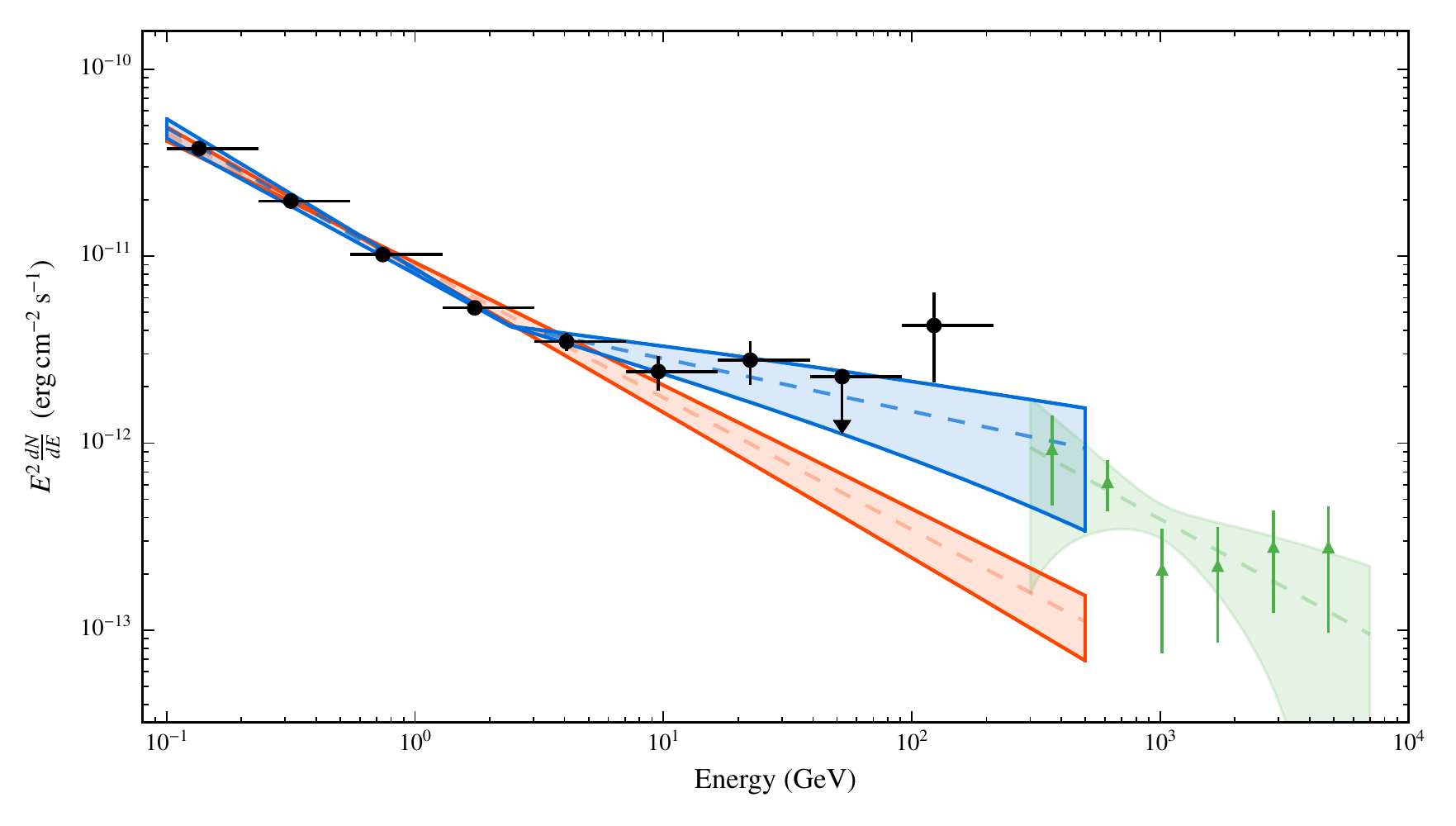}
\caption{A combined spectrum of our LAT analysis, shown with black circles, and the H.E.S.S. spectrum above $E_{\gamma}=250$~GeV, taken from \cite{HESSCenA2009}, shown in red. The broken power-law model, shown in blue, is preferred over the power law model, shown in red, with a significance $>5\sigma$. The grey histogram, with the right hand y-axis, shows the TS value for each spectral bin for the LAT data points. The bin with a TS$<25$, is replaced with an upper limit at 95$\%$ confidence level. Note that the last upper limit is not shown since it overlaps in energy with the H.E.S.S. data points. The broken power law fit allows for a smooth transition with the $\gamma$-ray spectrum reported here and the H.E.S.S. spectrum.}
\label{combinedspectrum}
\end{figure*}

Our analysis shows a hardening of the $\gamma$-ray spectrum above $2.6$ GeV at a significance level of 5$\sigma$. Importantly, while an extrapolation of the power-law description of the \textit{Fermi}-LAT spectrum to very high energies would under-predict the flux observed by the H.E.S.S. telescope array by an order of magnitude, the spectral hardening above $2.6$ GeV allows us to reconcile our \textit{Fermi}-LAT spectrum with previous H.E.S.S. observations. Furthermore, the non-variable flux we observe with \textit{Fermi}-LAT above the spectral break is also in agreement with the H.E.S.S. observations of Cen A finding no evidence of variability in the $E_{\gamma}>250$ GeV flux. The combined \textit{Fermi}-LAT and H.E.S.S. spectrum is shown in Figure \ref{combinedspectrum}.

With the existence of the hardening feature confirmed, we now turn our attention to understanding the origin of said feature. Traditionally, the $\gamma$-ray emission from radio galaxies has been attributed to a single zone population of relativistic electrons within the radio galaxy's jet that up-scatters the synchrotron photon field via the inverse Compton process \cite{latcore}, usually cited as the synchrotron self-Compton (SSC) model. In the SSC model, the high-energy tail of the electron population is responsible for both the X-ray and $\gamma$-ray emission and as such, if the SSC model was an accurate description of Cen A's core, the spectral break we observe in the $\gamma$-ray spectrum, should also be present in the X-ray spectrum, which is not seen in the latest X-ray observations \cite{Fuerst2016}. Furthermore, the lack of variability above the spectral break, combined with the evidence for variability below the spectral break, is difficult to reconcile with a single-zone SSC model as high energy electrons are expected to cool quicker than the low energy electrons, thus generating flux variability above the break energy rather than below.

A number of possible explanations for a spectral hardening, combined with non-variable emission above 2.6 GeV, are mentioned in \cite{Rieger2012}. One such model is a pulsar-like magnetospheric acceleration mechanism of electrons, with the entire MeV-GeV $\gamma$-ray flux attributed to inverse-Compton emission from the same lepton population \cite{Neronov2007}. However, this is inconsistent with both the spectral break and the lack of flux variability above 2.6 GeV. Moreover these magnetospheric models predict that inverse-Compton luminosity, $L_{\rm{IC}}$, is proportional to the super-massive black hole mass, $M_{\rm{BH}}$, which is not observed across the mass range of prominent $\gamma$-ray bright radio galaxies \cite{Brown2011}. 

Another possible explanation is that the hardening feature is associated with the decay of the pion by-products caused by a hadronic population within the jet interacting with an ambient photon field (eg. \cite{anita1,anita2}), as was done by \cite{Petropoulou} and more recently by \cite{cerruti}. However, given the lack of variability above the spectral break, this explanation would require the pion population to be in a `steady-state'.

Building on the apparently smooth transition between the data points above 2.6 GeV derived in this work and the H.E.S.S. observations, \cite{HESSCenA2009}, we discuss the validity of models that can account for the spectral hardening, by jointly fitting these data points. More specifically, we use the first three H.E.S.S. points, but exclude the last three which are less statistically significant, and more likely to be modified with new observations and an updated analysis. In any event, the last three points do not affect the best-fit spectrum due to their large error bars, but simply lead to a slight increase in the $\chi^{2}$.

\subsection{Dark Matter}
Dark matter (DM), as yet undetected, is a central explanation for structure formation, the stability of galaxies and the acoustic peaks of the Cosmic Microwave Background \cite{PlanckCollaboration2014a}. Most DM models assume  annihilation into lighter (Standard Model) particles but the nature of DM is still an open question. While it is established that DM agglomerates in the cores of galaxies, its density profile near the central black hole (BH) has yet to be characterized. Regardless, DM annihilations should produce high energy cosmic rays in the central parts of galaxies. The decay or hadronization of the particles  injected by the DM is expected to produce $\gamma$-rays at an energy smaller than or equal to the DM mass (assuming they are charged and unstable). Here we consider only prompt emission. We disregard any acceleration process \cite{Davis2015} or secondary emission \cite{Spike_GC_my_paper,M87_limits_my_paper} that could distort the prompt spectrum.  Here we examine whether this prompt emission is sufficient to explain the anomalous spectral component between 2.6 GeV and $\sim5$~TeV.

Throughout this work, we have assumed the existence of a spike in the distribution of the DM---induced by the adiabatic growth of a supermassive black hole at the centre of the galaxy---with a power-law index of $\gamma_{\mathrm{sp}} = 7/3$. Such a spiky profile has been suggested in Ref.~\cite{spikeGS}, although it is very debated. Indeed, a plethora of astrophysical phenomena could lead to a much shallower profile, such as non-adiabatic BH growth (as expected if the BH seed was brought in by a merger) \cite{Ullio2001,Gnedin2004} or an off-center position of the central BH, while others may have the opposite effect of making the formation or survival of spikes more likely, such as for instance enhanced accretion of DM to counteract the depopulation of chaotic orbits in a triaxial DM halo \citep{Merritt2004triaxial}.

It is also important to remember that even if a spike could form, the process of dynamical relaxation by DM scattering off stars could smooth down the spike and lead to a DM halo profile of the form $\rho \propto r^{-3/2}$ instead \cite{Merritt2004triaxial,Gnedin2004}. However, Cen A is dynamically young: its relaxation time is estimated to be $t_{\mathrm{r}} \sim 10^{2}$ Gyr (compared to $\sim 2.5$ Gyr for the Milky Way) due to the dependence on the mass of the central black hole. Hence we will assume that a spike formed in the core of Cen A at early times has survived dynamical processes.

\begin{figure}[h!]
\begin{center}
\includegraphics[width=1.0\linewidth]{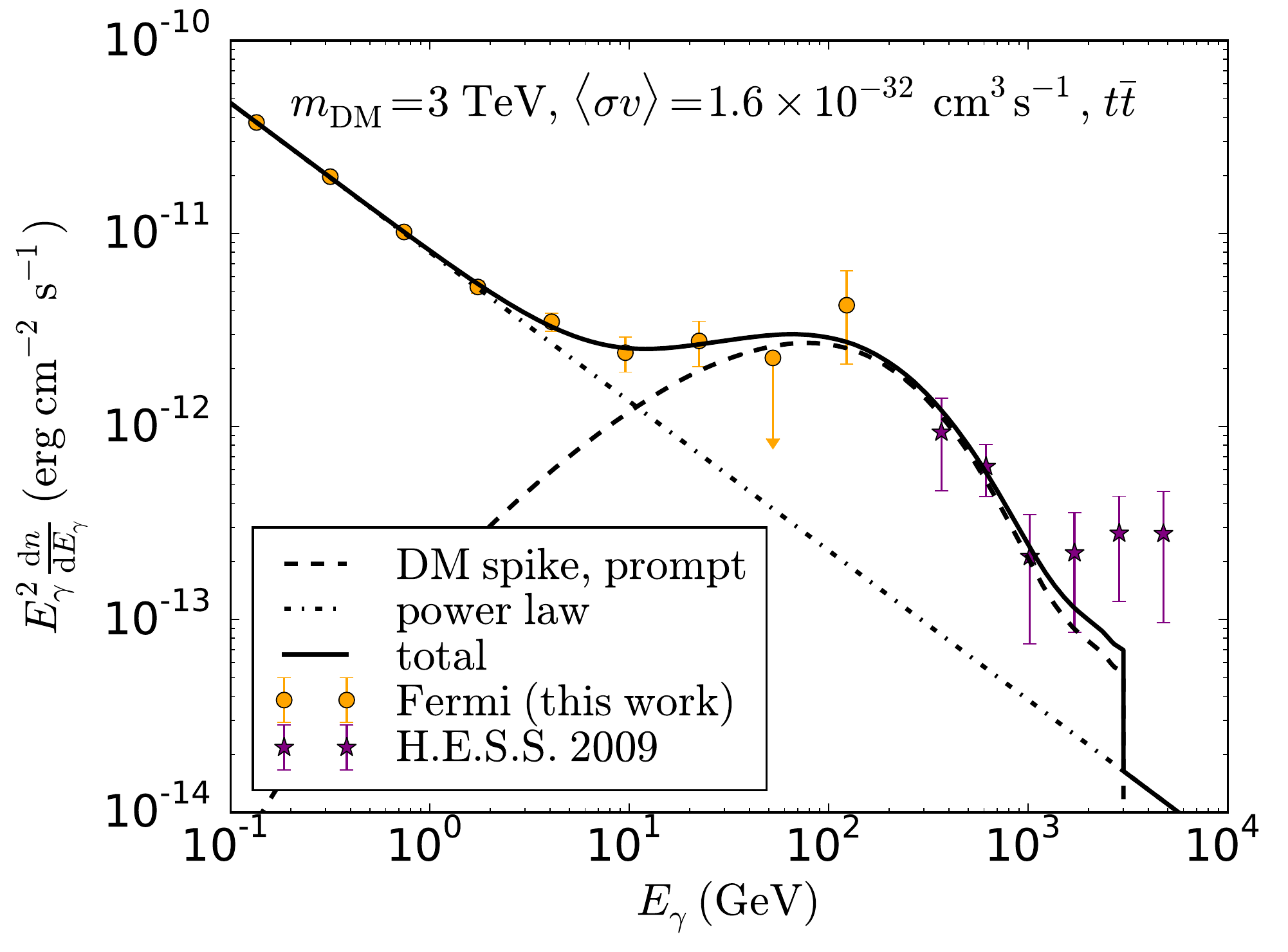}
\caption{\label{spike_t_tbar}Best fit to the $\gamma$-ray spectrum of Cen A, obtained by assuming a single power law plus the prompt emission from 3 TeV DM particles annihilating into $t\bar{t}$ with a cross-section of $1.6 \times 10^{-32}\ \mathrm{cm^{3}\ s^{-1}}$, and a spike in the DM density profile, with slope $\gamma_{\mathrm{sp}} = 7/3$.}
\end{center}
\end{figure}

In our best-fit analysis,  we consider DM self-annihilations into leptons or quarks and leave the DM mass and value of the annihilation cross-section (as well as the normalization and slope of the spectral power law emission below 2.6 GeV) free to vary. As shown in Fig.~\ref{spike_t_tbar}, our best fit favours a DM candidate with a mass of 3 TeV, annihilating into a pair of top and anti-top quarks ($t \bar{t}$) with cross-section  $<\sigma v> \simeq 1.6 \times 10^{-32} \ \rm{cm^3 s^{-1}}$, and a spike in the density profile. While the impact of a spike on the $\gamma$-ray emission from DM prompt emission in an AGN was first studied in \cite{M87_limits_my_paper}, our work constitutes the first evidence that $\gamma$-ray observations have the means to probe an anomalously high concentration of DM in the very inner core of AGNs. 

\begin{figure}[h!]
\centering
\includegraphics[width=1.0\linewidth]{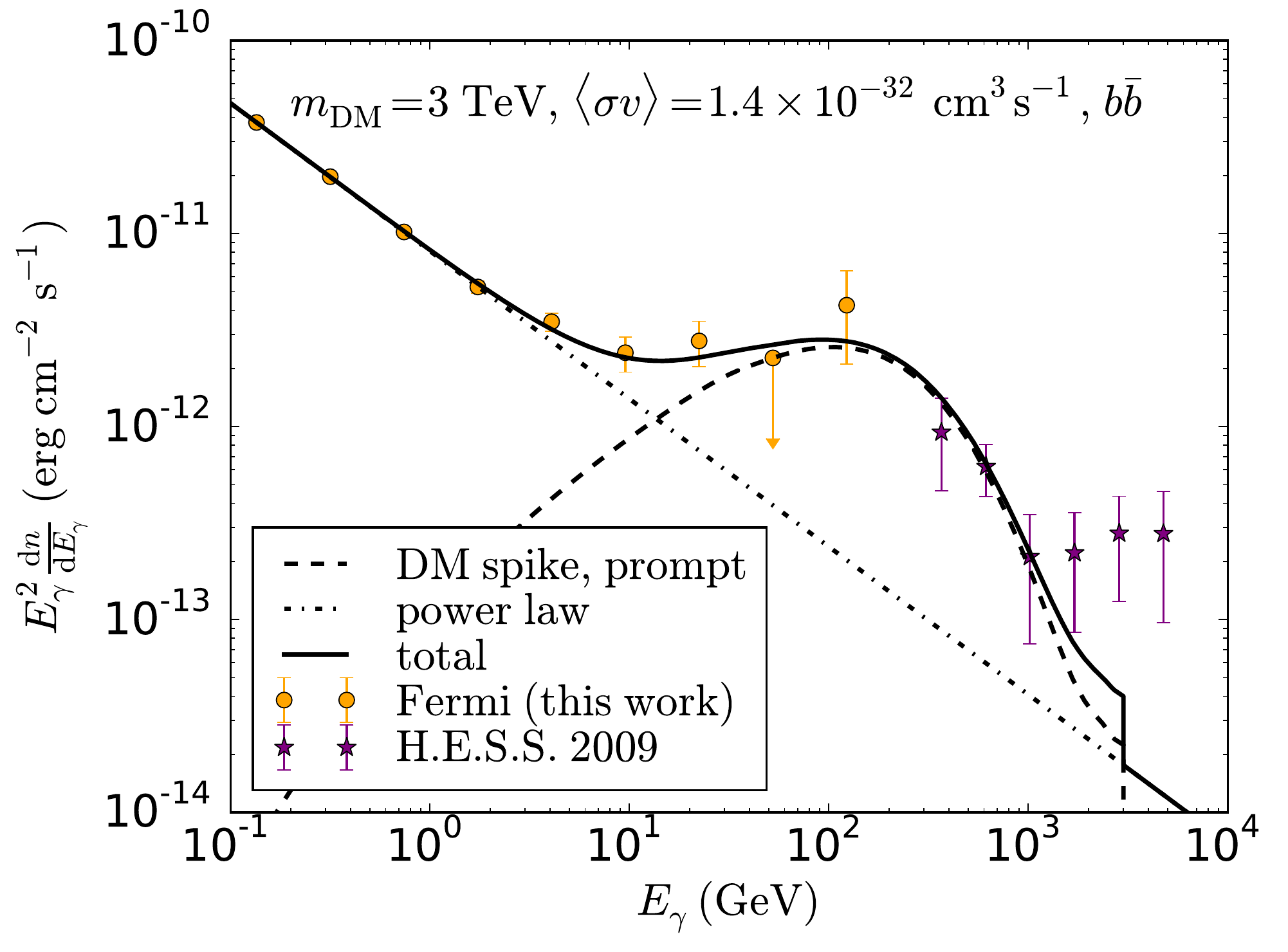} 
\caption{\label{spike_b_bbar} Best fit to the $\gamma$-ray spectrum of Cen A, obtained by assuming a single power law plus the prompt emission from 3 TeV DM particles annihilating into $b\bar{b}$ with a cross-section of $1.4 \times 10^{-32}\ \mathrm{cm^{3}\ s^{-1}}$, and a spike in the DM density profile, with slope $\gamma_{\mathrm{sp}} = 7/3$.}
\end{figure}

For a DM spike with slope $\gamma_{\mathrm{sp}} = 7/3$, the best fit shown in Fig.~\ref{spike_t_tbar}, and corresponding to annihilations into $t \bar{t}$ gives $\chi^2 = 1.7$, for 11 spectral data points. The points are taken from our \textit{Fermi}-LAT analysis and the first three data points from H.E.S.S. observations \cite{HESSCenA2009} and 4 free parameters ($m_{\rm{DM}}$, $<\sigma v>$ and the normalization and the slope of the power law spectrum below 2.6 GeV). We find  a $\chi^2/\rm{d.o.f} = 0.24$ which illustrates a remarkably good fit and suggests that it is dominated by statistical errors. Annihilations into $b\bar{b}$ give $\chi^2 = 3$, i.e.~$\chi^2/\rm{d.o.f} = 0.43$, which also corresponds to a very good fit (see Fig.~\ref{spike_b_bbar}). If we include the last three H.E.S.S.~points, we obtain $\chi^{2}/\mathrm{d.o.f.} = 0.62$ for $t\bar{t}$ and $\chi^{2}/\mathrm{d.o.f.} = 0.87$ for $b\bar{b}$.

\begin{figure}[h!]
\centering
\includegraphics[width=1.0\linewidth]{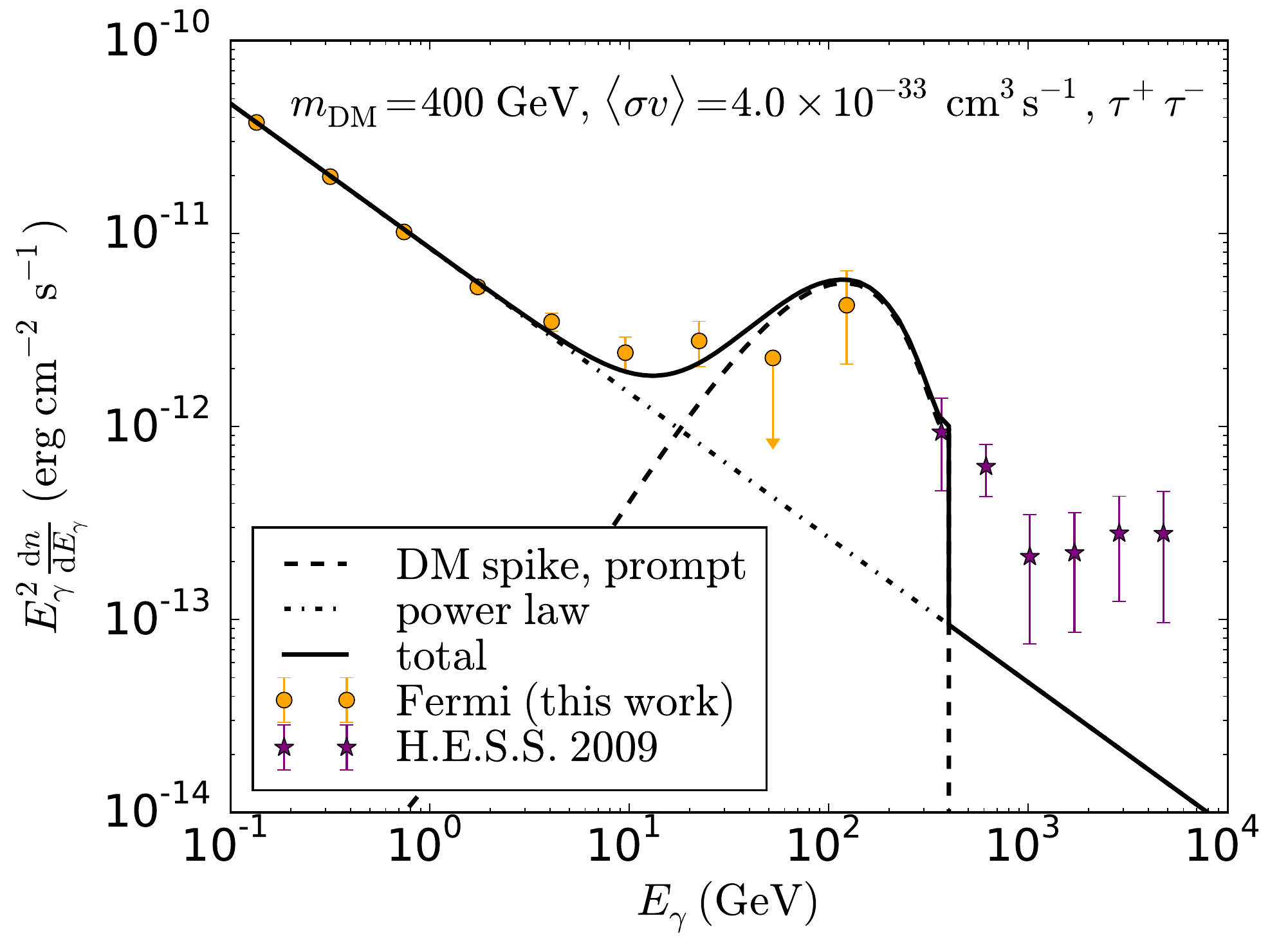} 
\caption{\label{spike_tau}Best fit to the $\gamma$-ray spectrum of Cen A, obtained by assuming a single power law plus the prompt emission from 400 GeV DM particles annihilating into $\tau^+\tau^-$ with a cross-section of $4 \times 10^{-33}\ \mathrm{cm^{3}\ s^{-1}}$, and a spike in the DM density profile, with slope $\gamma_{\mathrm{sp}} = 7/3$.}
\end{figure}

In Fig.~\ref{spike_tau}, we show the best fit when considering DM particles annihilating into $\tau^+\tau^-$. Now the best-fit mass is $\sim 400\ \mathrm{GeV}$, smaller than for hadronic channels. In this case, we find $\chi^{2}/\mathrm{d.o.f.} = 2.14$, so this channel gives a less good fit than the previous cases, but it still reproduces the spectral hardening reported in this work. For simplicity we have focused only on single annihilation channels; depending on the underlying model, DM particles may annihilate into several different final states.

\begin{figure}[h!]
\centering
\includegraphics[width=1.0\linewidth]{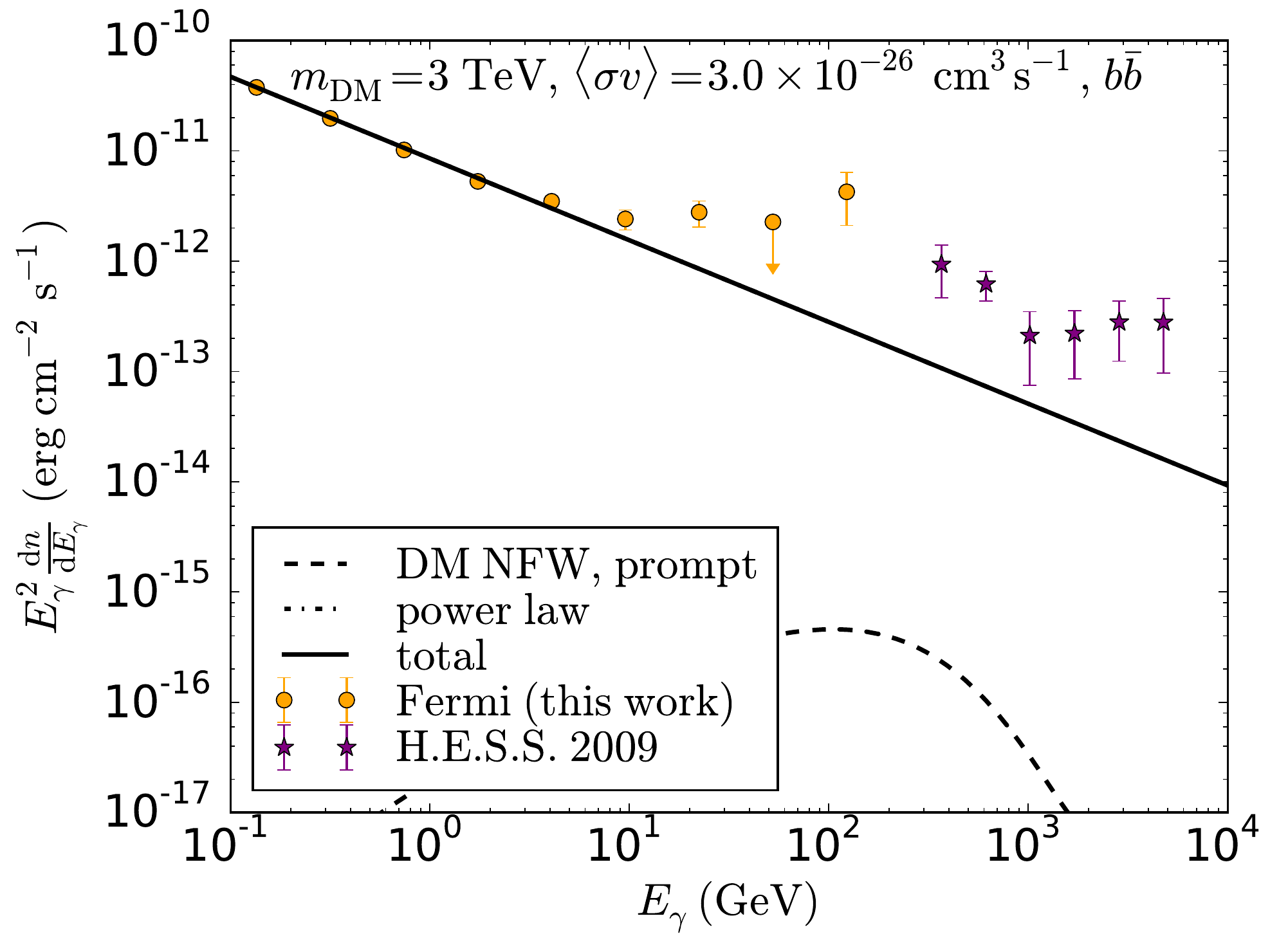} 
\caption{\label{NFW_b_bbar}Contribution to the $\gamma$-ray spectrum of Cen A from a single power law plus the prompt emission from 3 TeV DM particles annihilating into $b\bar{b}$ with the canonical cross-section of $3 \times 10^{-26}\ \mathrm{cm^{3}\ s^{-1}}$, for a DM density following the NFW profile. The DM-induced emission is several orders of magnitude below the data.}
\end{figure}

In the absence of a spike, DM annihilation for a maximum cross-section of $ <\sigma v> \simeq 3 \times 10^{-26} \ \rm{cm^3\ s^{-1}}$ cannot account for the spectral hardening, since in that case the DM flux is several orders of magnitude smaller than the observed flux. This is illustrated in Fig.~\ref{NFW_b_bbar}, for the NFW profile. In practice a spike with slope $\gtrsim 2$ is needed to account for the observed flux.

Our best-fit annihilation cross-section of the order of  $<\sigma v> \simeq 1.6 \times 10^{-32} \ \rm{cm^3 s^{-1}}$ is far too small to explain the observed fraction of DM in the Universe. However, this might simply be revealing the existence of a rich dark sector with several (non-thermal) DM particles \cite{BoehmFayetSilk2004,Zurek2009}\footnote{Our estimate assumes a unique DM candidate. Sub-component DM particles might require efficient co-annihilation processes, see e.g. \cite{Boehm2000}.}, or  a combination of velocity-dependent and independent terms in the annihilation cross section. We note that there is a degeneracy between the annihilation cross-section and the spike characteristics (normalisation, size, power law index) which could affect our estimates. Either way, our findings would suggest the existence of heavy DM particles exchanging at least one  heavy particle mediator.

%

\subsection{Millisecond pulsars}

An alternative explanation to the DM model is a population of millisecond pulsars (MSPs). These objects are rapidly rotating neutron stars which have been observed mostly in globular clusters. They are believed to spin up to millisecond periods due to frequent interactions with neighboring stars (e.g. by angular momentum accretion from a binary companion). For this very reason, they are  expected to form in high stellar density environments, including the central parsec around the Milky Way Galactic center (GC), where the density is considerably higher than in globular clusters. 

\begin{figure}[h!]
\begin{center}
\includegraphics[width=\linewidth]{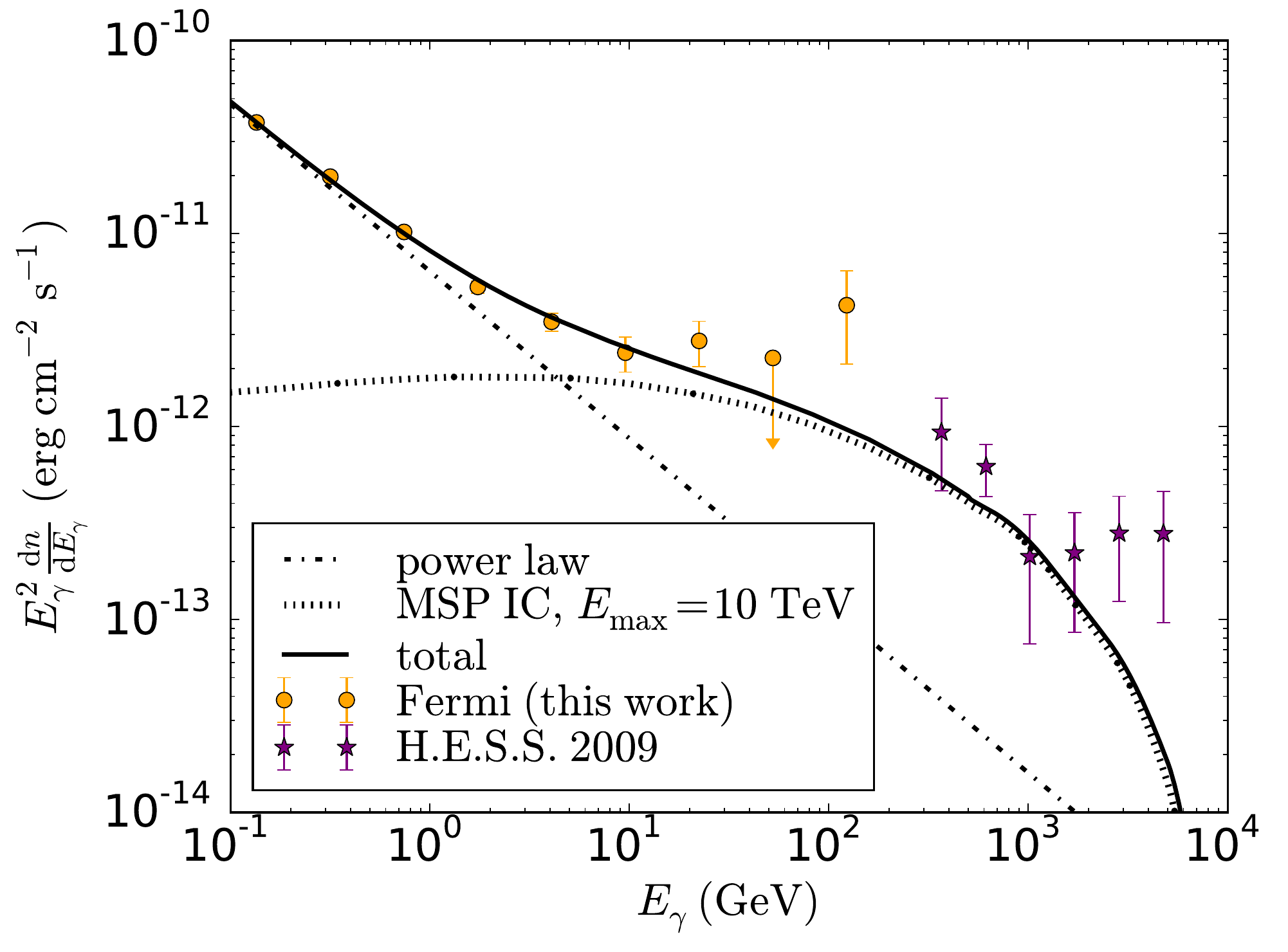}
\caption{\label{MSPs}Best fit for a population of MSPs plus a power law.}
\end{center}
\end{figure}

MSPs, along with DM, are currently the leading explanations for the low energy $\gamma$-ray excess which has been observed in the central region of the Milky Way (see e.g.~Ref.~\cite{Goodenough2009}). It is estimated that around $10^{3}$--$10^{4}$ MSPs are needed to explain the GeV excess in the GC~\cite{Petrovic2015}.  The production mechanism of high energy $\gamma$-rays from MSPs involves electron-positron pairs that may eventually be accelerated, radiate and produce more electron-positron pairs \cite{Harding2007}. This should lead to a spectral feature at GeV energies. However the resulting $\gamma$-ray signature can be broadened up to TeV energies by inverse Compton processes \cite{Bednarek2013}, when the electrons accelerated by MSP winds up-scatter the ambient soft photon field (from e.g. UV and IR bands). Two critical assumptions for these estimates are an electron injection spectrum extending to a few tens of  TeV and a large enough interstellar radiation field  for the inverse Compton losses to dominate over synchrotron losses. Here we use the same propagation technique as for the Milky Way (see e.g.~Ref.~\cite{GeV_excess_my_paper}). For the magnetic field, we assume a constant value of 10 $\mu \rm{G}$  constrained by \textit{Chandra} limits on synchrotron radiation from high energy electron-positron pairs and use a conservative power law, $\rho_{\rm{MSP}} \propto r^{-2.4}$, for the MSP density profile (consistent with the GC $\gamma$-ray excess). As for the DM analysis, we keep the normalization and slope of the jet contribution as free parameters.

\begin{figure}[h!]
\begin{center}
\includegraphics[width=\linewidth]{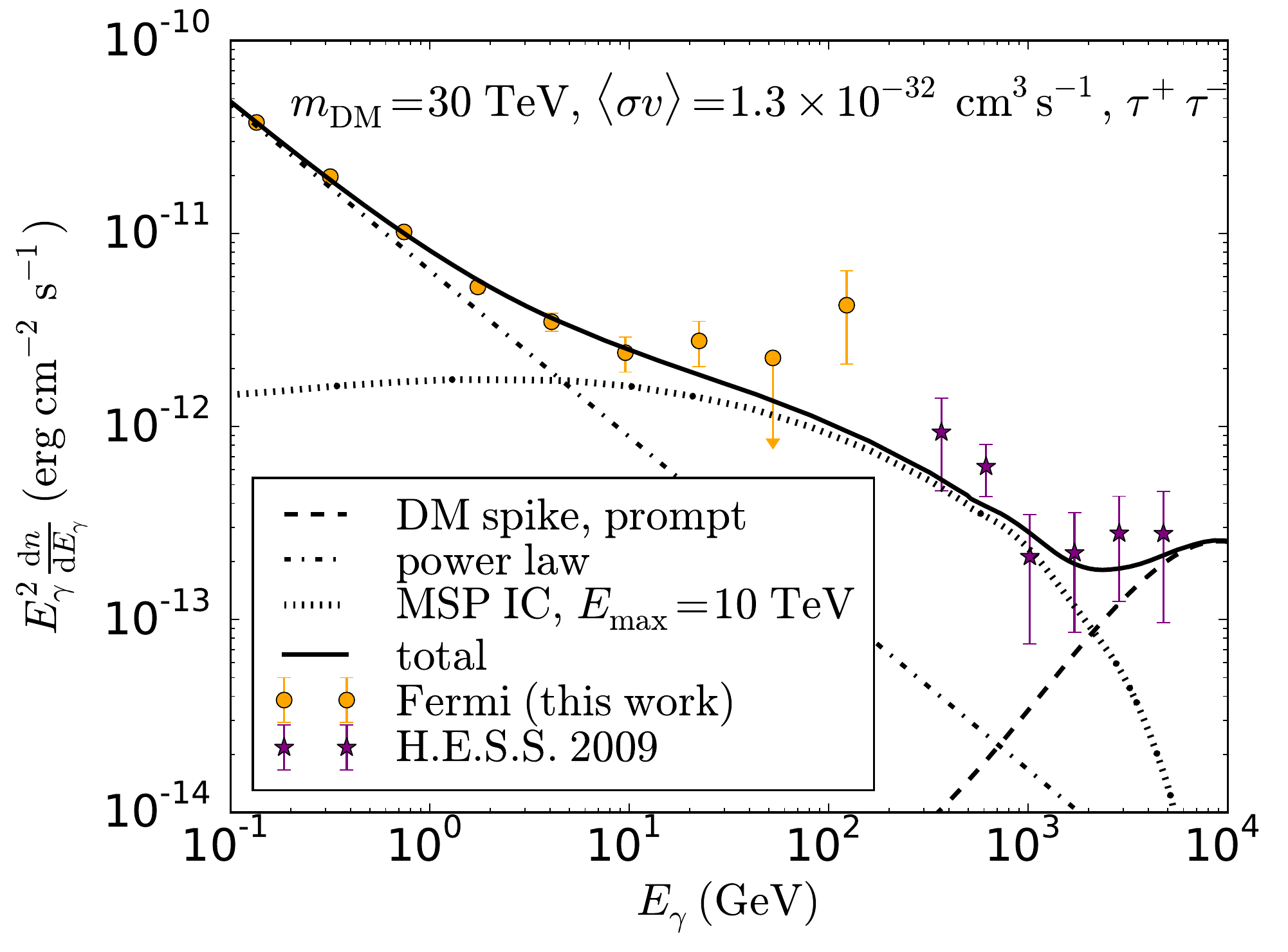}
\caption{\label{MSPs+spike}Best fit for a power law, a population of MSPs and a DM candidate of 30 TeV plus a spike.}
\end{center}
\end{figure}

Assuming the existence of such a population of MSPs leads to the best fit shown in Fig.~\ref{MSPs}, corresponding to $\chi^2 \simeq 10$, that is $\chi^2/\rm{d.o.f} \simeq 1.4$. This fit improves with the addition of a DM component at very high energy, as shown in Fig.~\ref{MSPs+spike} where we assumed a 30 TeV DM candidate annihilating into tau leptons. While the reduced $\chi^2$ is good, the poor knowledge of the MSPs density profile, the soft-photon field and our rather crude model of the magnetic field severely limits our interpretation of the goodness of the fit.


\section{Conclusion  \label{sec:conclusion}}
In this paper, we report a  5 $\sigma$  evidence for a hardening of the \textit{Fermi}-LAT $\gamma$-ray spectrum and show that either heavy DM particles  or a population of MSPs could explain this high-energy spectral feature. While we cannot rule out that the jet itself is  at the origin of the  hardening of the spectrum, the lack of variability of the emission above 2.6 GeV, both within the \textit{Fermi}-LAT  and H.E.S.S. spectra, argues against jet-induced leptonic models (such as SSC). Hence, at the very least, our results are a strong indication that the modelling of  $\gamma$-ray production mechanisms in active galaxies needs to be modified. 

Our findings hint at new physics inside objects like Cen A or astrophysical objects that are rarely detected outside our Galaxy. Therefore the precise modelling of cosmic ray propagation, together with observations of Cen A's $\gamma$-ray spectrum at TeV energies by H.E.S.S. and the forthcoming Cherenkov Telescope Array\cite{cta}, will be critical for determining the origin of the spectral break.   

A DM explanation would reveal the existence of particles beyond the standard model of particle physics as well as spikes in the DM profile. While the existence of spikes is highly debated, Cen A is dynamically young, so such an enhancement may survive nuclear star cluster dynamics. The fit to the combination of \textit{Fermi} and H.E.S.S. data in Figure 10 could provide tantalizing first evidence for the clustering of heavy dark matter particles around
black holes. If confirmed by other observations, our case for a boost of the DM annihilation signal constitutes the first evidence that DM clusters around black holes, and have important implications for our understanding of the feedback mechanisms between ordinary matter and DM in galactic cores. 

Likewise, if the spectral hardening of Cen A's core reported here is due to a population of MSPs, then the $\gamma$-ray spectrum of Cen A constitutes the first insights into the pulsar population in another galaxy. Either way, these findings open up a new window on the physics of the cores of active galaxies and provide the prospect of  more exciting discoveries.


\begin{thebibliography}{99}

\bibitem[\protect\citeauthoryear{Abdo et al.}{2010a}]{latcore} Abdo, A.A., et al., 2010a, ApJ, 719, 1433
\bibitem[\protect\citeauthoryear{Abdo et al.}{2010b}]{latlobe} Abdo, A.A., et al., 2010b, Science, 328, 725
\bibitem[\protect\citeauthoryear{Aarsten et al.}{2014a}]{3fgl} Acero,  F., et al., 2015, ApJS, 218, 23 
\bibitem[\protect\citeauthoryear{Aarsten et al.}{2014a}]{cta} Acharya, B.S., et al., 2013, Astropart. Physics, 43, 3
\bibitem[\protect\citeauthoryear{Ackermann et al.}{2012}]{lat2} Ackermann, M., et al., 2012, ApJS, 203, 4
\bibitem[\protect\citeauthoryear{Aarsten et al.}{2014a}]{Aharonian2006b} Aharonian, F.A., et al., 2006, Science, 314, 1424
\bibitem[\protect\citeauthoryear{Aarsten et al.}{2014a}]{HESSCenA2009} Aharonian, F.A., et al., 2009, ApJL, L40, 695
\bibitem[\protect\citeauthoryear{Aarsten et al.}{2014a}]{lat} Atwood, W.B., et al., 2009, ApJ, 697, 1071

\bibitem[\protect\citeauthoryear{Aarsten et al.}{2014a}]{Bednarek2013} Bednarek,W. \& Sobczak,T., 2013, MNRAS, 435, L14

\bibitem[\protect\citeauthoryear{boehm et al.}{2004}]{BoehmFayetSilk2004} B{\oe}hm, C., Fayet, P., \& Silk, J., 2004, Phys. Rev. D, 69, 101302, 10

\bibitem[\protect\citeauthoryear{boehm et al.}{2004}]{Boehm2000} B{\oe}hm, C., 2000, Phys. Rev. D, 62, 035012, 3


\bibitem[\protect\citeauthoryear{Brown \& Adams}{2011}]{Brown2011} Brown, A.M. \& Adams, J., 2011, MNRAS, 2785, 413
\bibitem[\protect\citeauthoryear{Aarsten et al.}{2014a}]{cerruti} Cerruti, M., Zech, A., Emergy, G \& Guarin, D., 2016, Proc. of the 6th International Symposium on High-Energy Gamma-Ray Astronomy, arXiv:1610.00255
\bibitem[\protect\citeauthoryear{Davis et al.}{2015}]{Davis2015} Davis, J.H., Silk, J., B{\oe}hm, C. and Kotera, K., 2016, Phys. Rev. D., 93, 3523

\bibitem[\protect\citeauthoryear{Evans \& Koratkar}{2004}]{evans} Evans, I.N \& Koratkar, A.P., 2004, ApJ, 617, 206
\bibitem[\protect\citeauthoryear{Aarsten et al.}{2014a}]{fr} Fanaroff, B.L. \& Riley, J.M., 1974, MNRAS, 167, 31
\bibitem[\protect\citeauthoryear{Aarsten et al.}{2014a}]{ferr} Ferrarese, L., et al., 2007, ApJ, 654, 186
\bibitem[\protect\citeauthoryear{Aarsten et al.}{2014a}]{Fuerst2016} Fuerst, F., et al., 2016, ApJ, 819, 150

\bibitem[\protect\citeauthoryear{gondolo and silk.}{2004}]{Goodenough2009} Goodenough, L. \& Hooper, D., 2009, arXiv:0910.2998

\bibitem[\protect\citeauthoryear{gondolo and silk.}{2004}]{Gnedin2004} Gnedin, O.Y., \& Primack, J.R.., 2004, Phys. Rev. Lett., 93, 061302
\bibitem[\protect\citeauthoryear{gondolo and silk.}{1999}]{spikeGS} Gondolo, P., \& Silk, J., 2003, Phys. Rev. Lett., 83, 1719


\bibitem[\protect\citeauthoryear{Hardcastle et al.}{2003}]{hardcastle} Hardcastle, M.J., et al, 2003, ApJ, 593, 169

\bibitem[\protect\citeauthoryear{Hardcastle et al.}{2003}]{Harding2007} Harding, A.K., Grenier, I.A. \&Gonthier, P.L., 2007, Astrophysics and Space Science, 309, 221

\bibitem[\protect\citeauthoryear{Jourdain et al.}{1993}]{granat} Jourdain, E., et al., 1993, ApJ, 412, 586
\bibitem[\protect\citeauthoryear{Lacroix et al.}{2014}]{Spike_GC_my_paper} Lacroix, T. and {B{\oe}hm}, C. and {Silk}, J., 2014a, Phys. Rev. D, 6, 063534, 89

\bibitem[\protect\citeauthoryear{Lacroix et al.}{2014}]{GeV_excess_my_paper} Lacroix, T. and {B{\oe}hm}, C. and {Silk}, J., 2014b, Phys. Rev. D, 90, 043508, 4

\bibitem[\protect\citeauthoryear{Lacroix et al.}{2015}]{M87_limits_my_paper} Lacroix, T. and {B{\oe}hm}, C. and {Silk}, J., 2015, Phys. Rev. D, 92, 043510, 4


\bibitem[\protect\citeauthoryear{Mattox et al.}{1996}]{like} Mattox, J.R., et al., 1996, ApJ, 461, 396

\bibitem[\protect\citeauthoryear{Merritt et al.}{1996}]{Merritt2004triaxial} Merritt, D. \& Poon, M.Y., 2004, ApJ, 606, 788
\bibitem[\protect\citeauthoryear{M\"{u}cke et al.}{2000a}]{anita1} M\"{u}cke, A., et al. 2000a, CPC, 124, 290
\bibitem[\protect\citeauthoryear{M\"{u}cke et al.}{2000b}]{anita2} M\"{u}cke, A., et al. 2000b, NuPhS, 80, 810

\bibitem[\protect\citeauthoryear{Aarsten et al.}{2014a}]{Neronov2007} Neronov, A. \& Aharonian, F.A., 2007, ApJ, 671,85
\bibitem[\protect\citeauthoryear{Aarsten et al.}{2014a}]{2fgl} Nolan,  P.L., et al., 2012, ApJS, 199, 31


\bibitem[\protect\citeauthoryear{Petropoulou et al.}{2014}]{Petropoulou} Petropoulou, M., Lefa, E., Dimitrakoudis, S. \& Mastichiadis, A., 2014, A\&A, 562, 12

\bibitem[\protect\citeauthoryear{Aarsten et al.}{2014a}]{Petrovic2015} Petrovi{\'c}, J., Serpico, P.D., Zaharijas, G., 2015, JCAP, 2, 23



\bibitem[\protect\citeauthoryear{Plank et al.}{2014a}]{PlanckCollaboration2014a} Planck Collaboration, 2014, A\&A, 571, 15

\bibitem[\protect\citeauthoryear{Aarsten et al.}{2014a}]{Rieger2012} Rieger, F., 2012, Memorie della Societa Astronomica Italiana, 127, 83
\bibitem[\protect\citeauthoryear{Aarsten et al.}{2014a}]{saha2013} Sahakyan, N., Yang, R., Aharonian, F.A. \& Reiger, F., 2013, ApJL, L6, 770
\bibitem[\protect\citeauthoryear{Aarsten et al.}{2014a}]{Sreekumar1999} Sreekumar, P., et al., 1999, Astroparticle Physics, 221, 11
\bibitem[\protect\citeauthoryear{Aarsten et al.}{2014a}]{steinle} Steinle,H., et al., 1998, A\&A, 330, 97
\bibitem[\protect\citeauthoryear{von Ballmoos et al.}{1987}]{balloons} von Ballmoos, P., et al., 1987, ApJ, 312, 134

\bibitem[\protect\citeauthoryear{Lacroix et al.}{2015}]{Ullio2001} Ullio, P., Zhao, H. \& Kamionkowski, M., 2001, Phys. Rev. D, 4, 043504, 64

\bibitem[\protect\citeauthoryear{Aarsten et al.}{2014a}]{urry} Urry, C.M \& Padovani, P, 1995, PASP, 107, 803

\bibitem[\protect\citeauthoryear{Lacroix et al.}{2015}]{Zurek2009} Zurek, K.M., 2009, Phys. Rev. D, 79, 115002, 11


\end{thebibliography}
\end{document}